\documentclass[10pt,conference]{IEEEtran}
\IEEEoverridecommandlockouts
\usepackage{cite}
\usepackage{amsmath,amssymb,amsfonts}
\usepackage{algorithmic}
\usepackage{graphicx}
\usepackage{textcomp}
\usepackage{xcolor}
\usepackage{enumitem} 
\usepackage{ifthen}
\usepackage{float}
\usepackage{comment}
\usepackage{url}
\usepackage{balance}
\usepackage{footnote}
\makesavenoteenv{tabular}
\makesavenoteenv{table}

\newboolean{showcomments}
\setboolean{showcomments}{true}
\ifthenelse{\boolean{showcomments}}
{ \newcommand{\mynote}[2]{\textcolor{orange}{
			\fbox{\bfseries\sffamily\scriptsize#1}
			{\small$\blacktriangleright$\textsf{\emph{#2}}$\blacktriangleleft$}}}}
{ \newcommand{\mynote}[2]{}}

\ifthenelse{\boolean{showcomments}}
{ \newcommand{\dnote}[2]{\textcolor{red}{
			\fbox{\bfseries\sffamily\scriptsize#1}
			{\small$\blacktriangleright$\textsf{\emph{#2}}$\blacktriangleleft$}}}}
{ \newcommand{\dnote}[2]{}}

\ifthenelse{\boolean{showcomments}}
{ \newcommand{\hnote}[2]{\textcolor{blue}{
			\fbox{\bfseries\sffamily\scriptsize#1}
			{\small$\blacktriangleright$\textsf{\emph{#2}}$\blacktriangleleft$}}}}
{ \newcommand{\hnote}[2]{}}

\def\BibTeX{{\rm B\kern-.05em{\sc i\kern-.025em b}\kern-.08em
    T\kern-.1667em\lower.7ex\hbox{E}\kern-.125emX}}
\begin{document}

\title{\textit{NICHE}: A Curated Dataset of Engineered Machine Learning Projects in Python
}

\author{
\IEEEauthorblockN{
Ratnadira Widyasari,
Zhou Yang,
Ferdian Thung,
Sheng Qin Sim, 
Fiona Wee,
Camellia Lok, 
Jack Phan, 
}
\IEEEauthorblockN{
Haodi Qi,
Constance Tan,
Qijin Tay, 
and David Lo
}

\IEEEauthorblockA{\textit{School of Computing and Information System, Singapore Management University}\\
\{ratnadiraw.2020,zyang,ferdianthung,sqsim.2018,fiona.wee.2018,camellialok.2017,jack.phan.2018\}@smu.edu.sg\\
\{haodi.qi.2017,hytan.2018,qijin.tay.2018,davidlo\}@smu.edu.sg}
}

\maketitle

\begin{abstract}
Machine learning (ML) has gained much attention and been incorporated into our daily lives. While there are numerous publicly available ML projects on open source platforms such as GitHub, there have been limited attempts in filtering those projects to curate ML projects of high quality. The limited availability of such high-quality dataset poses an obstacle in understanding ML projects. To help clear this obstacle, we present \textit{NICHE}, a manually labelled dataset consisting of 572 ML projects. Based on evidences of good software engineering practices, we label 441 of these projects as engineered and 131 as non-engineered. This dataset can help researchers understand the practices that are followed in high-quality ML projects. It can also be used as a benchmark for classifiers designed to identify engineered ML projects.

\end{abstract}

\section{Introduction}
There are many valuable pieces of information stored in a version control system of a project; they include: source code, documentation, issue reports, test cases, list of contributors, etc. Researchers mine these software repositories to get useful insights related to how bugs are fixed \cite{bug_fix}, how developers collaborate \cite{network} and so on. With the abundance of the open source repositories in GitHub, researchers can mine for insights and validate hypotheses on a large corpus of data. However, Kalliamvakou et al. showed that most repositories in Github are of low-quality \cite{Kalliamvakou2014, Kalliamvakou2015AnIS}, which can lead to wrong and biased conclusions. To avoid skewed findings, researchers usually take some measures to filter out low-quality projects, e.g., by choosing projects with a high number of stars (which is considered to reflect the projects' popularity). Unfortunately, popularity may not be correlated with project quality \cite{Sajnani2014}. Therefore, Munaiah et al. propose an approach to find high-quality software projects, more specifically; by identifying engineered software projects \cite{Munaiah2017}. Such projects are essential for mining software repository (MSR) research, as they allow for high-quality findings to be uncovered (from high-quality data).

Machine learning (ML) projects are becoming increasingly popular and play essential roles in various domain, e.g., code processing~\cite{PTM4TAG,9825884}, self-driving cars, speech recognition \cite{6638947}, etc. Despite widespread usage and popularity, only a few research works try to examine AI and ML projects to identify unique properties, development patterns, and trends. Gonzalez et al. \cite{10.1145/3379597.3387473} find that the AI \& ML community has unique characteristics that should be considered in future software engineering and MSR research. For example, more support is needed for Python as the main programming language, and there are significant differences between internal and external contributors in AI \& ML projects.
We coin a term for such research: Mining Machine Learning Repository (MLR).
Similar to conventional MSR research, MLR also requires high-quality projects. In GitHub, there are many tutorials, resource pages, courseworks and toy projects that are related to ML; some of which are very popular but unsuitable for MLR research. 
To facilitate MLR research, we present a curated dataset of E\textbf{N}g\textbf{I}neered Ma\textbf{CH}ine L\textbf{E}arning Projects in Python (\textit{NICHE}). We first automatically identify projects in GitHub that: (1) use one of the popular ML libraries, and (2) satisfy some basic quantitative quality metrics. This process returns 572 ML projects from GitHub. Next, we manually analyze the 572 ML projects and label them as engineered or not engineered. This dataset can be used as the raw material for MLR research, or as the benchmark for evaluating classifiers designed to identify engineered ML projects.

We label the dataset manually to ensure high quality and accurate labels. Our criteria for assessing an ML project are rooted in Munaiah et al. work~\cite{Munaiah2017}. We check 8 distinct dimensions of a project (architecture, community, CI, documentation, history, issues, license and unit testing) to evaluate whether the project is engineered or not. 
Out of the 572 projects we collected, 441 projects are labelled as engineered ML projects, and 131 projects are labelled as non-engineered ML projects. There are several related datasets in the literature. Datasets from \cite{Munaiah2017} and \cite{PHANTOM} have labels indicating whether a project is engineered or not, but they do not contain ML projects. Gonzalez et al.\cite{10.1145/3379597.3387473} collected a dataset of ML \& AI projects, but these projects are not comprehensively assessed based on their adoption of good software engineering practices. 
They only eliminate tutorials, homework assignments and so on.
We make our dataset publicly available\footnote{https://doi.org/10.6084/m9.figshare.21967265}.

The rest of this paper is organized as follow. Section 2 describes the methodology used to collect and filter the dataset, as well as how the dataset is stored. Section 3 gives an overview of the dataset. In Section 4, we propose some potential research applications that the dataset can be used for. In Section 5, we discuss some related work. Section 6 briefly mentions threats to validity. Finally, we conclude the paper and present future work in Section 7.

\section{Dataset Collection Methodology}
\begin{table*}[!t]
\centering
\caption{Engineered ML Project in Python: OpenNMT/OpenNMT-tf (left) \& Non Engineered ML Project in Python: mesutpiskin/computer-vision-guide (right)}
\label{tab:example_engineered}
\begin{tabular}{p{1.5cm} p{7.3cm} p{7.7cm}} 
\hline
& {\raggedright\textbf{OpenNMT/OpenNMT-tf}} & \textbf{mesutpiskin/computer-vision-guide} \\
\hline
\textbf{Architecture}           & The source code of this project has been organized into different modules in a way that shows the relationship between different files. & The source code of this project has been organised into different folders.\\ \hline
\textbf{Community}              & 20 actives contributors found in https://github.com/OpenNMT/OpenNMT-tf/graphs/contributors. & There are no contributors to the project other than the author.     \\ \hline
\textbf{Continuous Integration} & The project uses Travis CI\footnote{https://travis-ci.org/} to hosted continuous integration service.                                & Continuous Integration (CI) is not used on the project.   \\ \hline
\textbf{Documentation}          & There is a well-maintained README, and the majority of the functions i.e.: runner.py, training.py, and text.py have appropriate docstrings.  & Most of the functions in the source code are not documented. Only have the general information regarding project on the README.md.  \\ \hline
\textbf{History}                & 20 commits in June 2020 as seen in their pulse monthly page out of total 1,735 commits from July 2017. & 2 commits in June 2020 as seen in their pulse monthly page out of total 101 commits from September 2018  
\\ \hline
\textbf{Issues}                 & The project maintainers respond to issues, as evidenced by issues that were quickly resolved by the developers\footnote{https://github.com/OpenNMT/OpenNMT-tf/issues/695}\footnote{https://github.com/OpenNMT/OpenNMT-tf/issues/698}.                                     & There is no issues raised or closed from May 2019. Only has one issues and the authors take more than 6 months to close this issue.       \\ \hline
\textbf{License}                & 
Project license is clearly written in their GitHub page.
& Project license is clearly written in their GitHub page.
\\ \hline
\textbf{Unit Testing}           & There are many files related to testing.                                                                                             & There is no evidence of any testing done on the project.      \\ \hline
                      
\end{tabular} 
\end{table*}

In this section, we discuss how we collect, curate and store the dataset.

\subsection{Automated Data Collection}
We use GitHub as our data source. GitHub is the most active open source community and project hosting site. It provides rich information related to projects, including popularity, pull requests, statistics of commit history, dependency graph, etc. These data enable us to easily determine whether a software project adopts good software engineering practices. 

This paper focuses on ML projects that use at least one of these popular ML libraries: Theano,\footnote{http://deeplearning.net/software/theano/} PyTorch,\footnote{https://pytorch.org/} and TensorFlow.\footnote{https://www.tensorflow.org/} GitHub provides dependency graphs which lists all the public projects that depend on a library. We fetch lists of projects from the 3 ML libraries' dependency graphs and also get some statistical information, e.g. number of stars, number of commits and timestamp of the latest commit. 
To filter out low-quality projects, we only consider projects that satisfy the following requirements: 
\begin{enumerate}
    \item \textbf{Popularity}: the project must have 100 or more stars, as we want to filter toy projects that are clearly non-engineered.
    \item \textbf{Activity}: the project must have at least one commit that is more recent than May 1 2020, as we want to analyze projects that are still updated when we collecting the data.
    \item \textbf{Project Availability}: the project should be public and accessible via GitHub API.
    \item \textbf{Commit History}: the project have at least 100 commits.  We use this as a threshold to filter out new projects or projects that only have a short history.
\end{enumerate}
After applying these filters, we have 572 projects for further analysis.

\subsection{Manual Data Curation}
 To ensure the quality of the dataset, we filter and label the ML projects manually. We adopt the framework used by Munaiah et al. \cite{Munaiah2017} to decide whether an ML project is an engineered project or not. In this framework, a software project is considered to be engineered if it there is a clear evidence of sound software engineering practices such as good architecture, good documentation, sufficient testing, evidence of proper project and so on. Here we list the eight dimensions Munaiah et al. \cite{Munaiah2017} used to assess a software project:

\begin{itemize}
    \item \textbf{Architecture}: Well-defined relationships between components within a project is an indicator of good code organization.
    \item \textbf{Community}: Presence of many collaborators in a project shows that there is a form of collaboration during the project development.
    \item \textbf{Continuous Integration (CI)}: Using CI ensures that the code is stable for development or release, which indicates that the project has a good code quality.
    \item \textbf{Documentation}: The presence of sufficient documentation shows developers' thinking of code maintainability.
    \item \textbf{History}: When the code have undergone continual changes or sustained evolution, it indicates that the project is being modified to ensure its viability.
    \item \textbf{Issues}: Projects that maintain management activities such as requirements, schedules, and tasks through GitHub Issues show proper project management.
    \item \textbf{License}: Having license in the project, explicitly shows the rights of the user in making copies of the project, which shows a good code accountability.
    \item \textbf{Unit testing}: Presence of testing in projects shows the developer's effort to ensure that the project work as it should, indicating that the project is of good quality.
\end{itemize}

There are no fixed criteria for deciding to what extent fulfilling these dimensions makes a project be engineered. But, in general, fulfilling one or two of these dimensions does not guarantee that the project is engineered. However, having the majority of the dimensions fulfilled makes the project likely to be engineered.


As the projects that we assess are still actively updated, it is possible that the status and label for these projects change over time. For our analysis, we collected the projects on June 2020. Each project is assessed by two authors who independently evaluate each dimension and give their final judgements. The authors check if the project satisfies the dimensions by looking at its code, documentation, contributors, issues, and other information available in GitHub.
The following are guidelines that authors use to assess a project: (1) look for evidence of sufficient testing, e.g. number of files related to test and code coverage (if applicable); (2) look for evidence that the project has architectural decisions in organizing the code, e.g. the code is split into different modules and no ad-hoc scripts; (3) check whether good documentations are provided; (4) check if the project uses issues to track new features and bugs; (5) check if the project uses a CI service, e.g. Travis, CircleCI, etc; (6) check if the project is updated within the last one month; (7) check how many active contributors the project has; (8) check whether the project provides a license. 

For every point in the guideline, we consider the following dimensions for the project assessment: unit testing for point (1), architecture for point (2), documentation for point (3), issues for point (4), CI for point (5), history for point (6), community for point (7), and license for point (8). Aside from providing a label on whether a project is engineered or not, the labellers also provide descriptive information for every dimension. In case both labellers do not agree (on whether a project is engineered or not), the disagreement is resolved by a discussion. This discussion is led by the author that does not label the project and is more experienced.


An example of a project that is labelled as engineered is shown in Table~\ref{tab:example_engineered} on the left column. In this example, the \texttt{OpenNMT/OpenNMT-tf}\footnote{https://github.com/OpenNMT/OpenNMT-tf} project is labelled as an engineered ML project since it satisfies all the dimensions in the evaluation framework.
For example, there is evidence of quick response to issues, usage of CI service, sufficient testing and so on. Unlike the engineered ML project, the non-engineered ML project shown on the right column of Table I does not fulfil enough dimensions. The mesutpiskin/computer-vision-guide\footnote{https://github.com/mesutpiskin/computer-vision-guide} is labelled as a non-engineered ML project because most of the dimensions in this project are not satisfied. For example, this project does not have an active community contributor, as the only contributor is the author himself. The project also does not have any test cases or employ CI, does not have much documentation, and only has one issue since the creation of the project.
The only dimension that is satisfied by the project is license.

Labelling this dataset takes around 123 person-hours in total. Out of the initial 572 projects that we collected using GitHub API, 441 projects are labelled as engineered ML projects and 131 projects are labelled as non-engineered.

\subsection{Data Storage}

Our dataset is available at  \url{https://github.com/soarsmu/NICHE}. 
We provide a CSV file that contains the  list  of  the  project names along with  their labels and descriptive  information  for  every  dimension. We also provide several basic statistics for each project, such as the number of stars and commits.

\vspace{0.1cm}
\section{Dataset Overview}
In this section, we provide some basic statistics, including the number of stars, commits and lines of code of the 572 projects in our dataset.

\subsection{Number of Stars}
\begin{figure}[t]
	\begin{center}
    \includegraphics[width=0.85\columnwidth]{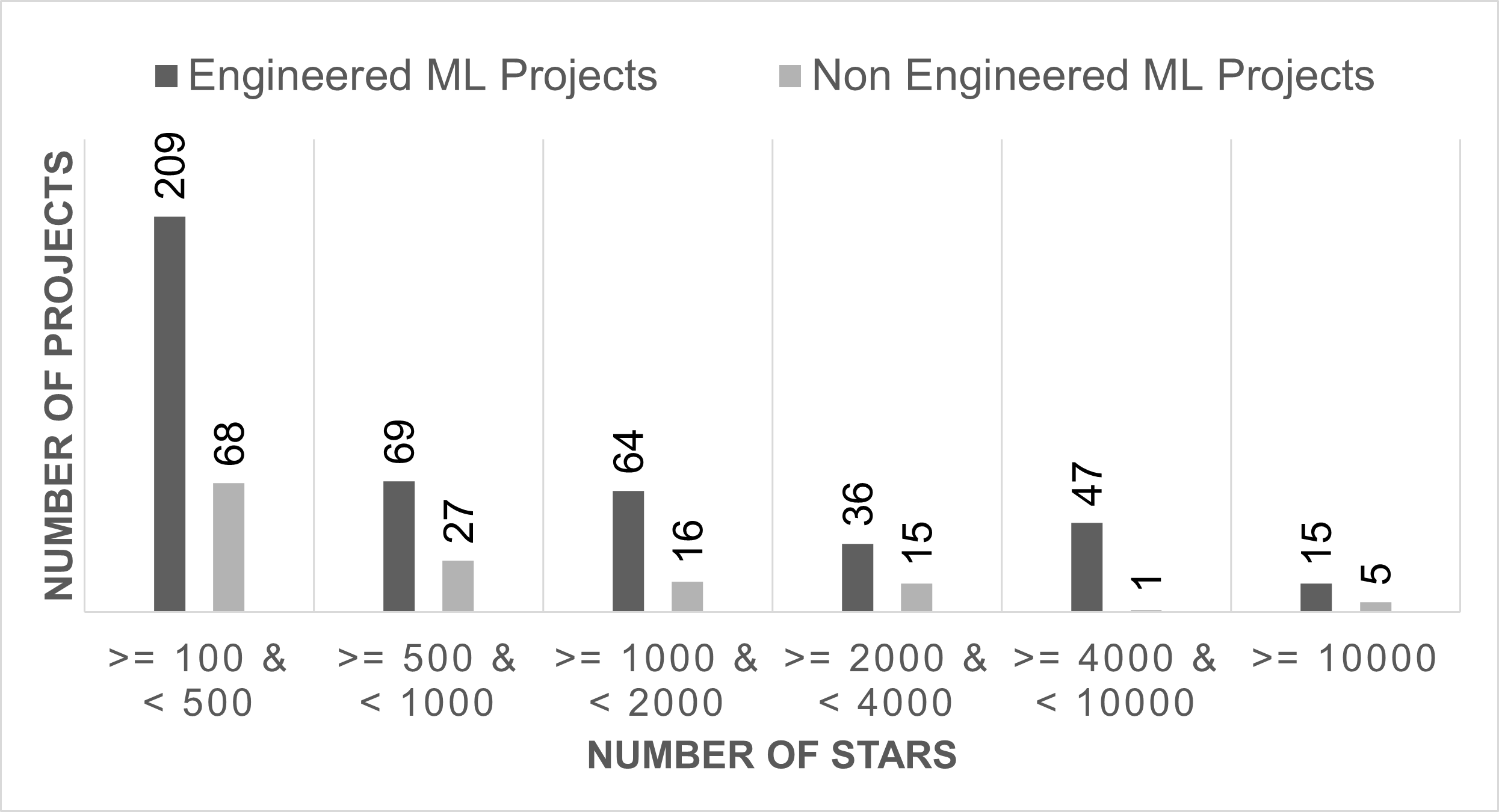} 
	\caption{Distribution of projects in terms of number of stars}
	\label{fig:stars}
	\end{center}
\end{figure}

The number of stars in GitHub reflects how popular the corresponding project is. The number of stars has been shown to positively correlate with the popularity of the project~\cite{star1, 7887704}. We count the number of stars for each project in our dataset.
In Figure~\ref{fig:stars}, we group projects based on their numbers of stars.
We observe that around half of the projects (280 out of 572) have between 100 and 500 stars. Around 75.45\% projects in this category are labelled as engineered ML projects. We also observe that projects with more stars are more likely to be engineered, but the inverse does not hold as only 47.5\% engineered projects have 100 to 500 stars. This observation is similar to previous findings by Munaiah et al.\cite{Munaiah2017}.


\subsection{Number of Commits}
\begin{figure}[t]
	
	\begin{center}
    \includegraphics[width=0.85\columnwidth]{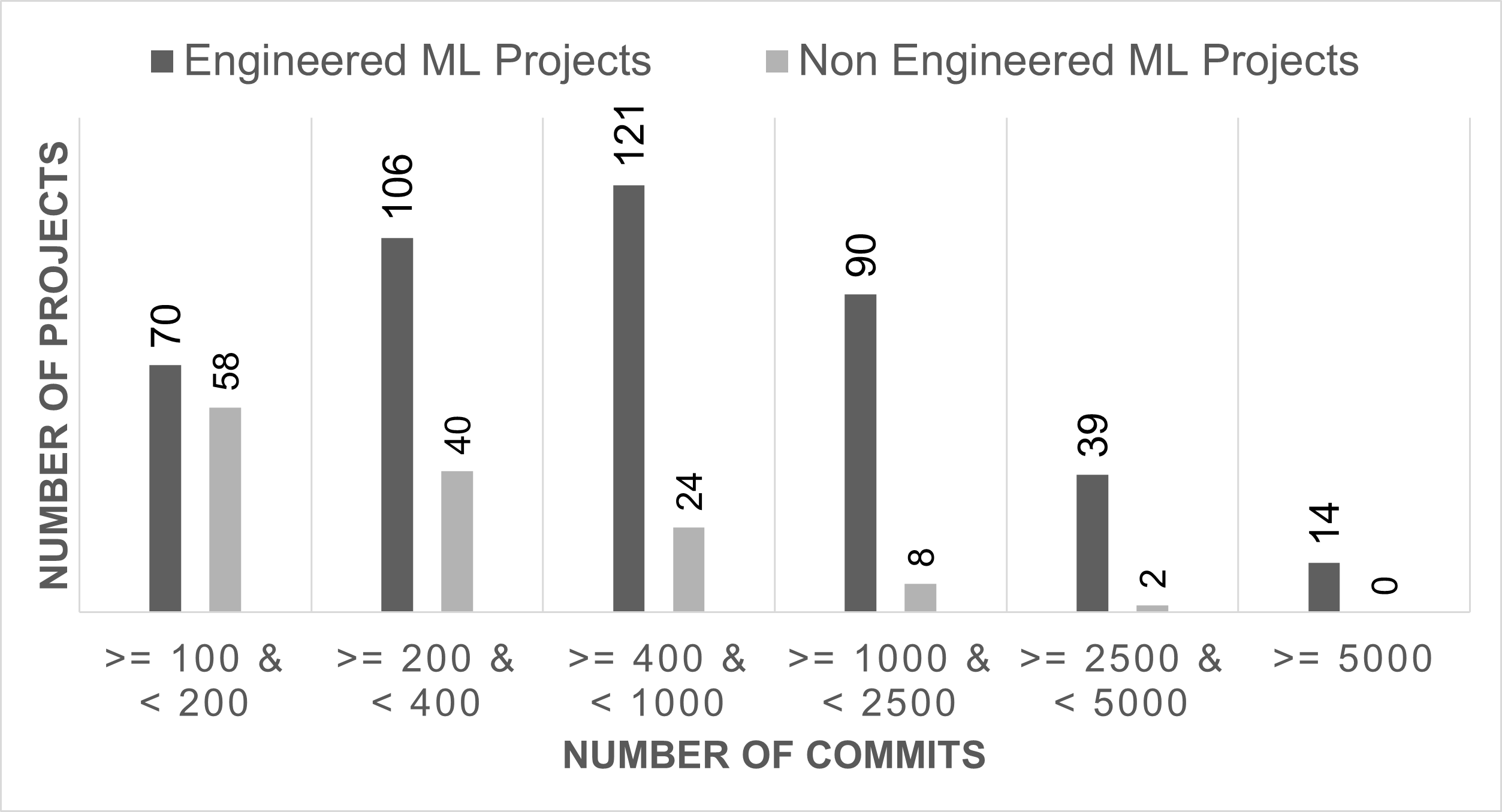} 
	\caption{Distribution of projects in terms of number of commits}
	\label{fig:commits}
    \end{center}

\end{figure}

The number of commits reflects a project development history. In Figure~\ref{fig:commits}, we group projects based on their numbers of commits.
One interesting pattern is that the more commits a project has, the more likely the project is an engineered ML project. 
This pattern is evidenced by the fact that the ratio of engineered ML projects increases as the number of commits increases.

\subsection{Lines of Code (LOC)}
\begin{figure}[t]
	
	\begin{center}
    \includegraphics[width=0.85\columnwidth]{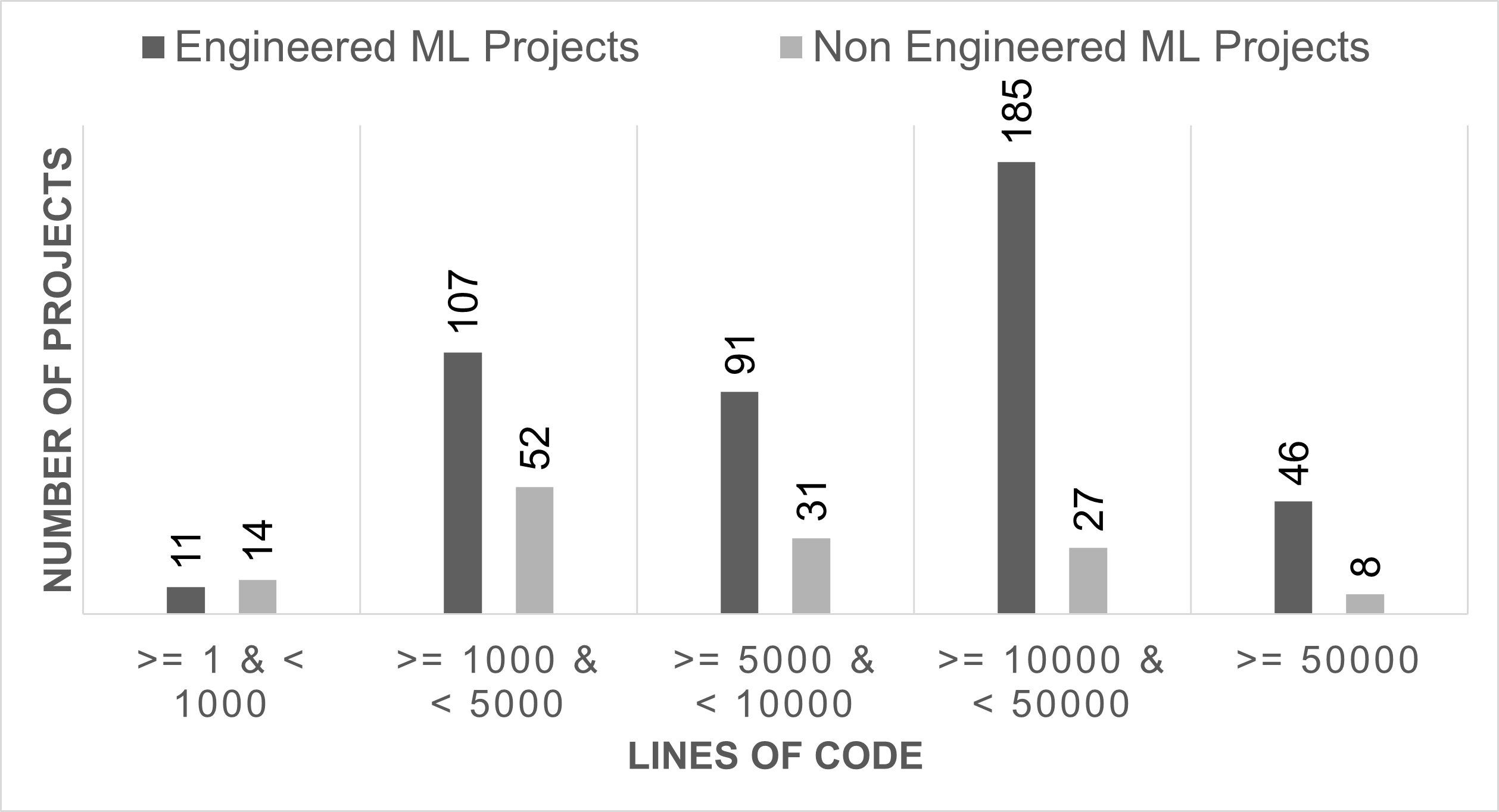} 
	\caption{Distribution of projects in terms of lines of code}
	\label{fig:sloc}
    \end{center}
\end{figure}

The number of Lines of Code (LOC) can be an indicator of project size. We clone all the collected projects and checkout their main branch. Then, we use cloc\footnote{https://github.com/AlDanial/cloc} to count LOC. We plot the distribution of projects' LOC in Figure~\ref{fig:sloc}. The largest project has 705,797 LOC and is labelled as engineered. For projects with less than 1,000 LOC, the number of non-engineered projects surpasses that of engineered ML projects. 
We observe that the ratio of engineered ML projects increases as the number of LOC increases.


\section{Potential Research Application}
Our dataset can be used to explore many research topics such as mining machine learning repositories and evaluating engineered ML project classifiers.

\noindent \textbf{Mining Machine Learning Repositories.} As machine learning projects become more popular and important, many researchers start to analyze properties of machine learning projects \cite{9240645, 10.1145/3379597.3387479, 10.1145/3379597.3387473,fan2021makes}, which we refer to as Mining Machine Learning Repository (MLR). High-quality ML projects are essential for MLR. Some research works were done on a few projects, e.g. Peng et al. find that Apollo's overall code coverage is not high \cite{ads}. Our dataset can enable researchers to validate a hypothesis at a larger scale. The dataset can be used to answer questions like how well engineered ML projects are covered by existing test cases, and what are the differences between the coverage of ML files and non-ML files in engineered ML projects.

\noindent \textbf{Benchmark for engineered ML project classifiers.} Reaper \cite{Munaiah2017} and PHANTOM \cite{PHANTOM} both show decent performance on classifying whether a software project is engineered or not. But the datasets that they are evaluated on contain only conventional software projects. There is no intersection between their dataset and ours. How their engineered software project classifiers perform on ML projects are unknown. The community may need classifiers specially designed for ML projects, and our dataset can be the benchmark for such classifiers.


\section{Related Work}

We identify three datasets that are most related to our work. One is the dataset from Munaiah et al. \cite{Munaiah2017}. 
Munaiah et al. provide a dataset of 800 manually labelled software projects. 400 of them are labelled as engineered and the other 400 are labelled as non-engineered. Another related dataset is from \cite{PHANTOM} which extends Munaiah et al.'s dataset \cite{Munaiah2017} with additional 200 projects. There is no overlap between their datasets and ours.
Gonzalez et al.\cite{10.1145/3379597.3387473} provide a list of ML-related projects. However, they only eliminate projects that are tutorials, homework assignments, coding challenges, ‘resource’ storages, or collections of model files/code samples. These projects have not been assessed based on the criteria of good software engineering practices that are considered by Munaiah et al.

\section{Threats to Validity}
To ensure the quality of our dataset, we manually curate ML projects. However, even though we have tried our best, it is still possible that we mislabel some projects. To mitigate potential mislabelling, the data is labelled independently by two authors. If there is any disagreement, a more experienced author leads a discussion to resolve the disagreement.


\section{Conclusion and Future Work}
To conclude, we present \textit{NICHE}, a dataset of ML projects in Python consisting of 441 projects that are identified as engineered ML projects and 131 projects that are identified as non-engineered ML projects. Our dataset contains the largest number of engineered ML projects in Python. The dataset is curated by hand to better ensure the correctness of the label assigned to each project. We believe the 441 engineered projects will make it easier for MSR researcher to get sizeable and high quality ML projects for their research, as it is very important to exclude poor quality projects. Moreover, the 131 non-engineered projects serve as negative examples that can be used to build a benchmark to build and evaluate a classifier that can distinguish engineered from non-engineered ML projects.

As future work, we also plan to investigate the properties of the engineered ML repositories to identify their unique characteristics, pain points faced by their contributors, and ways for researchers to help them.


\balance
\bibliographystyle{IEEEtran}
\bibliography{./Bibliography}

\end{document}